\documentclass[english]{article}
\usepackage[T1]{fontenc}
\usepackage[latin9]{inputenc}
\usepackage{amsmath}
\usepackage{amssymb}
\usepackage{cancel}
\usepackage{graphicx}
\usepackage{esint}

\makeatletter
\newcommand{\lyxaddress}[1]{
	\par {\raggedright #1
	\vspace{1.4em}
	\noindent\par}
}

\makeatother

\usepackage{babel}
\begin{document}
\title{Meson-Baryon Couplings Revisited}
\author{N. F. Nasrallah}
\maketitle

\lyxaddress{\begin{center}
Faculty of Science, Lebanese University. Tripoli 1300, Lebanon
\par\end{center}}
\begin{abstract}
The theoretical evaluation of the coupling constants $g_{\pi NN}$,
$g_{KN\Lambda}$and $g_{KN\Sigma}$ is undertaken using QCD sum rules.
These quantities were previously calculated with exponential (Borel)
kernels used to suppress the unknown contributions of the hadronic
continua. This method however introduces arbitrariness and instability
in the calculation. In order to avoid these I redo the calculation
using polynomial kernels tailored to vanish at the baryonic resonance
masses. The results are $g_{\pi NN}=11.5\pm1.0$, $g_{KN\Sigma}=5.45\pm.4$,
$g_{KN\Sigma}=-(12.7-15.0)$ which are close to experiment and to
the predictions of SU(3) and which , with the corresponding Goldberger-Treiman
Discrepancy satisfy quite well the Dashen-Weinstein relation.
\end{abstract}

\section{Introduction}

The $\pi$$N$ coupling constant $g_{\pi NN}$ is one of the most
important parameters in hadron physics. This quantity has been extensively
studied since the works of Reinders, Rubinstein and Yazaki \cite{RRY}
using the method of QCD sum rules \cite{SVZ} starting with either
the correlator of three interpolating fields or with the pion to vacuum
matrix element of two nucleon interpolating currents. These correlators
introduce several different Dirac structures and these were studied
in detail two decades ago \cite{S and H}, \cite{B and K}, \cite{Kim}
and more recently in \cite{Azizi}

All these calculations however involve dispersion integrals which
include the unknown contribution of the continuum with quantum numbers
of the nucleon. In order to minimize this unknown contribution a damping
kernel $e^{-\frac{t}{M^{2}}}$ is introduced. This method however
presents problems of arbitrariness and stability. I present here a
different choice for the damping kernel: a polynomial P(t) which practically
vanishes in the nucleon resonance region. The sum rule method consists
of expressing the quantity of interest, which enters in the residue
at the pole of the correlation function $\Pi(t)$ in terms of an integration
of $\Pi(t)$ over the contour $c$ shown in Fig.1 in the complex t-plane
$.$This in turn is the sum of an integral over the cut of the discontinuity
of $\varPi(t)$ and an integral over the circle of large radius R
over which $\Pi(t)$ can be replaced by its QCD expression $\Pi^{QCD}(t)$
the latter can be brought back to an integral over the cut of the
discontinuity of $\Pi^{QCD}(t).$ The main unknown in the calculation
is the integral over the cut of the hadronic amplitude.

In order to minimize this contribution an integration kernel P(t)
is introduced so that 
\begin{equation}
\text{Residue x }P(\text{pole})=\frac{1}{\pi}\int_{th}^{R}dtP(t)Im\Pi(t)+\frac{1}{\pi}\int_{0}^{R}dtP(t)Im\Pi^{QCD}(t)\label{eq:1.1}
\end{equation}

The usual choice for the kernel is $P(t)=e^{-\frac{t}{M^{2}}}$where
$M^{2}$ (the Borel mass) is a damping parameter which cannot be too
large because the damping worsens nor can it be too small because
the contribution of higher orders in the Operator Product Expansion
(OPE) of $\Pi^{QCD}(t)$ increase beyond control. An intermediate
value of $M^{^{2}}$has to be chosen by stability considerations which
is not always possible. The radius R of the circle is another adjustable
parameter.

In order to avoid these problems I shall choose for the kernel a simple
polynomial $P(t)=1-a_{1}t-a_{2}t^{2}$ tailored to practically vanish
in the nucleon resonance region in order to render the first integral
on the r.h.s. of eq. (\ref{eq:1.1}) negligible. The value of R is
chosen to lie in the stability region of the second integral on the
r.h.s. of eq. (\ref{eq:1.1}). This leads to an unambiguous result.

The method is applied to the calculation of the coupling constants
$g_{_{\pi NN}},$$g_{KN\varLambda}$and $g_{KN\Sigma}$.

\begin{figure}[h]
\includegraphics[width=0.6\textwidth]{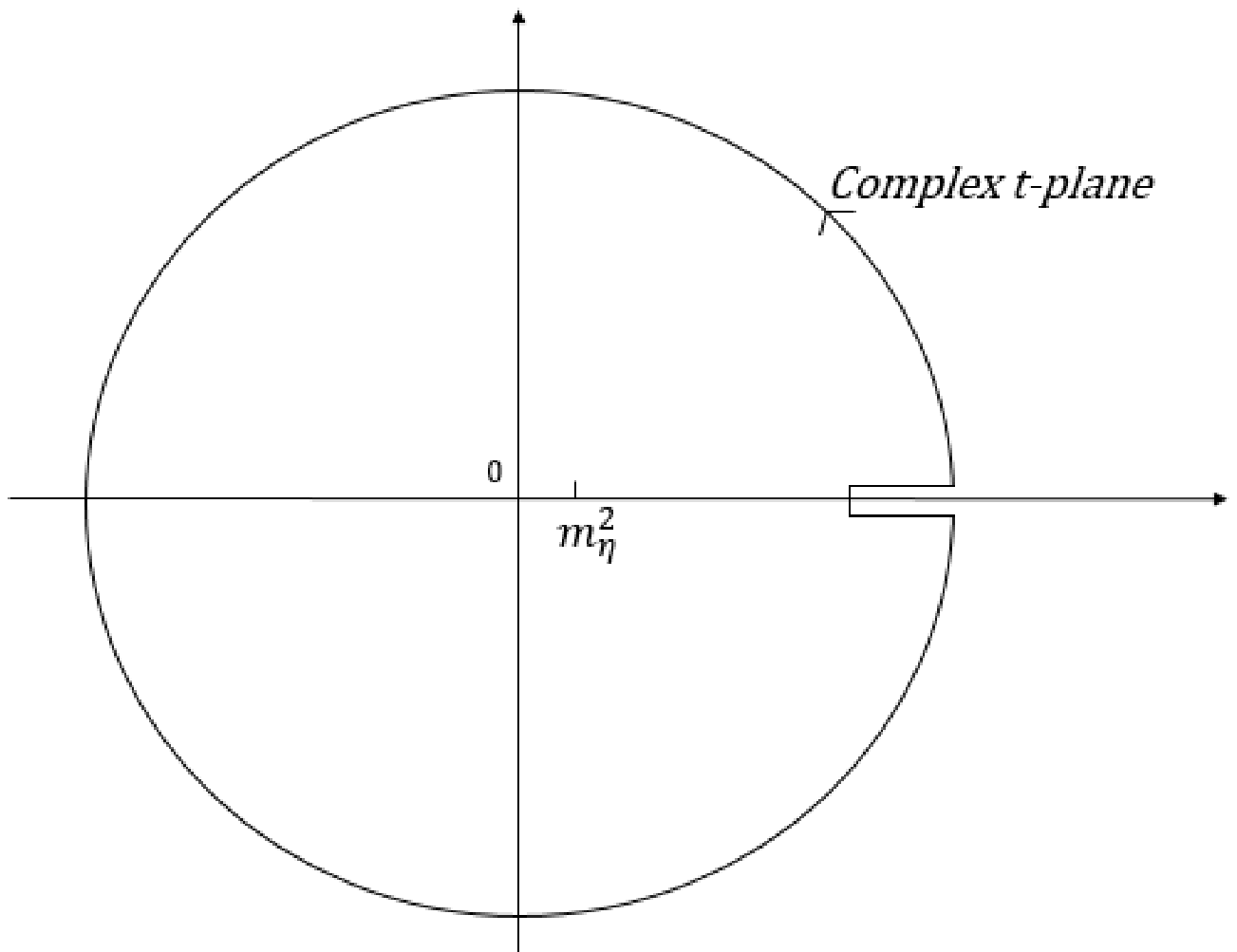}
\begin{centering}
Fig. 1
\par\end{centering}
\end{figure}

\section{The Pion-Nucleon Coupling Constant g$_{\pi NN}$}

Consider the correlation function

\begin{equation}
\Pi(p,q)=i\int dxe^{iqx}\left\langle 0\left|Tj_{p}(x)j_{n}(0)\right|\pi^{+}(p)\right\rangle \label{eq:2.1}
\end{equation}

where 
\begin{equation}
j_{p}=\epsilon_{abc}(u_{a}^{T}C\gamma_{\mu}u_{b})\gamma_{5}\gamma_{\mu}d_{c}\label{eq:2.2}
\end{equation}
is the proton interpolating field of Joffe \cite{Joffe}. The interpolating
field of the neutron is obtained by the interchange of $u$ and $d$.

I shall work with the $\gamma_{5}\sigma_{\mu\nu}q_{\mu}q_{\nu}$ structure
of the correlator.

In the soft pion limit the OPE reads \cite{Kim}: 

\begin{equation}
\Pi^{QCD}(t)=\sqrt{2}\left\langle qq\right\rangle \left[\frac{ln(-t)}{12\pi^{2}f_{\pi}}+\frac{4}{3}\frac{f_{\pi}}{t}-\frac{C}{t^{2}}\right]\label{eq:2.3}
\end{equation}
with $t=q^{2}$ ,
\begin{equation}
C\simeq\left(\frac{1}{216f_{\pi}}\left\langle a_{s}GG\right\rangle -\frac{m_{o}^{2}f_{\pi}}{6}\right)\simeq-.0117GeV^{3}\label{eq:2.4}
\end{equation}
at the standard values $\left\langle a_{s}GG\right\rangle \simeq.012GeV^{^{4}},m_{0}^{2}\simeq.8GeV^{2}$

On the hadronic side $\Pi(t)$ has a double nucleon pole, single nucleon
poles and a continuum.

\begin{equation}
\Pi(t)=\frac{\sqrt{2}g_{_{\pi NN}}\lambda_{N}^{2}}{(t-m_{N})^{2}}+\text{single poles + continuum}\label{eq:2.5}
\end{equation}

$\lambda_{N}$denotes the coupling of $j_{N}$to the nucleon 
\begin{equation}
\left\langle 0\left|j_{N}\right|N\right\rangle =\lambda_{N}U_{N}\label{eq:2.6}
\end{equation}

Consider the integral $\frac{1}{2\pi i}$$\int_{c}dt$P(t)$\Pi(t)$
in the complex t-plane where c is the contour shown in Fig.1.

As argued in the introduction I shall choose $P(t)=1-a_{1}t-a_{2}t^{2}$
with the coefficients $a_{1,2}$ tailored in order to practically
annihilate the contribution of the continuum. We thus have
\begin{equation}
\sqrt{2}g_{\pi NN}\lambda_{N}^{2}P'(m_{N}^{2})+\varDelta=\frac{1}{\pi}\int_{th}^{R}dtP(t)Im\Pi(t)\Pi+\frac{1}{2\pi i}\oint dtP(t)\Pi^{QCD}(t)\label{eq:2.7}
\end{equation}

$\varDelta$ is the contribution of the single poles. The choice of
$P(t)$ is based on the fact that the contribution of the first integral
on the r.h.s. of eq. (\ref{eq:2.6}) arises mostly from the interval
\cite{N S} $I=2.0GeV^{2}\preceq t\preceq3.0GeV^{2}$ where the resonances
$N^{+}(1440),N^{-}(1535),N^{-}(1650),N^{+}(1710)$ lie. The parameters
$a_{1}$and $a_{2}$ are chosen so as to minimize the integral $\int_{2.0GeV^{2}}^{3.0GeV^{2}}dt$
$\left|P(t)\right|^{2}$, numerically $a_{1}=.807GeV^{-2},a_{2}=-.160GeV^{-4}$. 

This allows the neglect of the first integral on the r.h.s. of eq.
(\ref{eq:2.7}) so that, using expression (\ref{eq:2.3}) for $\Pi^{QCD}(t)$

\begin{equation}
\sqrt{2}g_{\pi NN}\lambda_{N}^{2}P'(m_{N}^{2})+\varDelta\cong\sqrt{2}\left\langle qq\right\rangle \left[\frac{1}{12\pi^{2}f_{\pi}}I_{0}(R)+\frac{4}{3}f_{\pi}-a_{1}C\right]\label{eq:2.10}
\end{equation}

with the definition
\begin{equation}
I_{n}(R)=\int_{_{0}}^{R}dtt^{n}P(t)\label{eq:2.11}
\end{equation}
The fact that the contrbution of term $a_{1}C$ is small compared
to that of the preceding ones ($\sim5\%$) justifies the neglect of
the next unknown term proportional to $a_{2}$. This is to be contrasted
to the case of exponential damping where an inifinite number of unknown
higher order terms enter. Morever it can be seen from Fig. 2 that
the damping of the continuum is much better in our case.

\begin{figure}[h]
\includegraphics[width=0.7\textwidth]{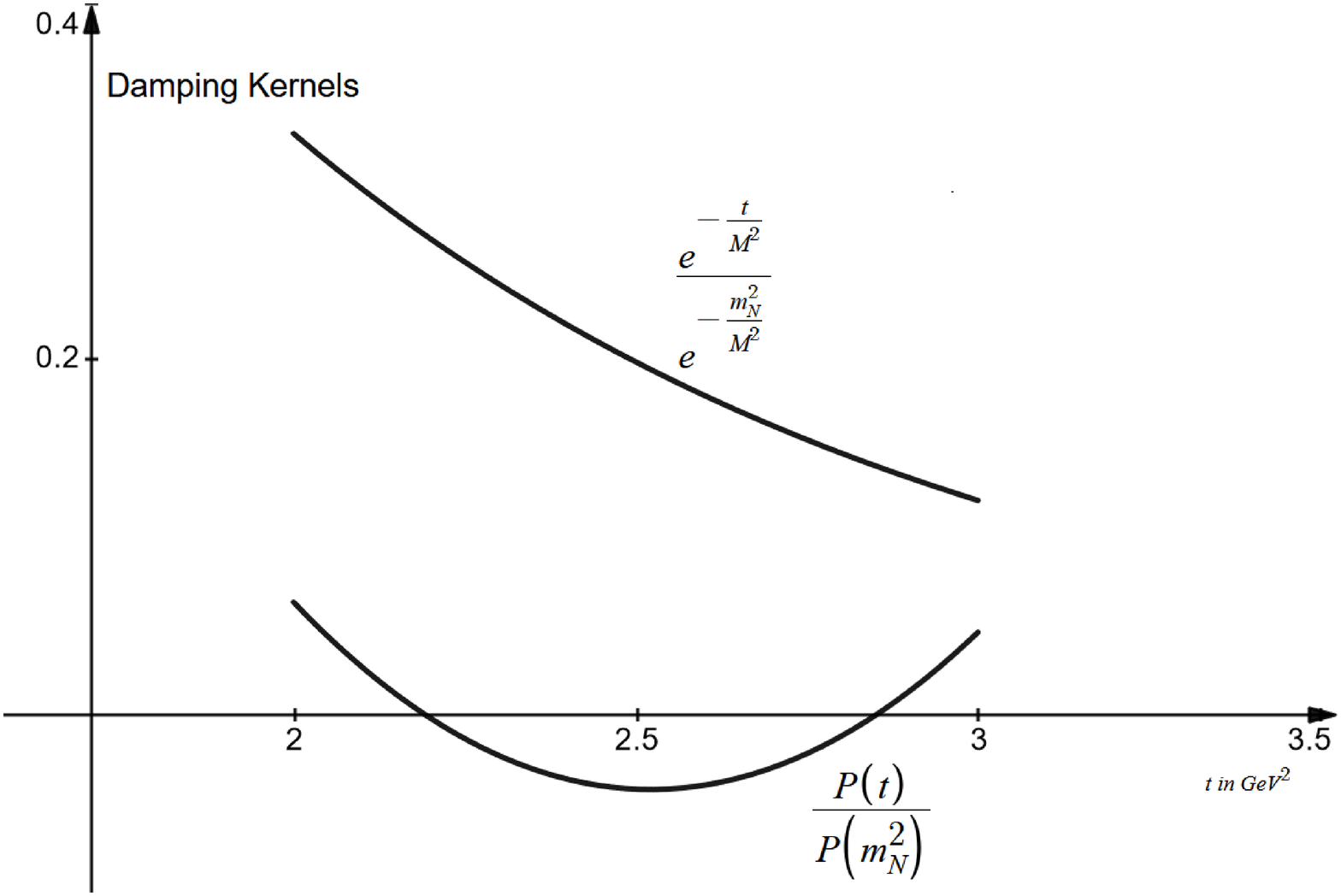}
\begin{centering}
Fig. 2 
\par\end{centering}
\begin{centering}
$P(t)$ is our polynomial $1-a_{1}t-a_{2}t^{2}$
\par\end{centering}
\centering{}$M^{2}\sim1.0GeV^{2}$as chosen in the litterature
\end{figure}

$\lambda_{N}$was obtained using the same approach in \cite{N S}
starting with the nucleon correlator
\begin{equation}
\Pi_{N}(q)=i\int dxe^{iqx}\left\langle 0\left|T\eta(x)\eta(0)\right|0\right\rangle =\cancel{q}\Pi_{1}(q^{2})+\Pi_{2}(q^{2})\label{eq:2.12}
\end{equation}

The sum rule using $\Pi_{2}(q^{2})$ yields

\begin{equation}
(2\pi)^{4}m_{N}\lambda_{N}^{2}P(m_{N}^{2})=-B_{3}I_{1}(R)-B_{7}\label{eq:2.13}
\end{equation}

with the OPE coefficients

\begin{equation}
B_{3}=4\pi^{2}(1+\frac{3}{2}a_{s})\left\langle \bar{q}q\right\rangle ,\,\,\,\,\,\,\,\,\,\,B_{7}=-\frac{4}{3}\pi^{2}\left\langle qq\right\rangle \left\langle a_{s}GG\right\rangle \label{eq:2.14}
\end{equation}

Let us now estimate $\varDelta$. The contribution of the $N'(1440)$
adds to eq. (\ref{eq:2.5})

\begin{equation}
\delta_{N'}=\sqrt{2}g_{\pi NN'}\lambda_{N}\lambda_{N'}\frac{1}{(t-m_{N}^{2})(t-m_{N'}^{2})}\label{eq:2.15}
\end{equation}

which corresponds to

\begin{equation}
\varDelta=\sqrt{2}g_{\pi NN}\lambda_{N}\lambda_{N'}\frac{P(m_{N}^{2})-P(m_{N'}^{2})}{m_{N}^{2}-m_{N'}^{2}}\label{eq:2.16}
\end{equation}

While $P(m_{N'}^{2})$ is small by construction $P(m_{N}^{^{2}})$
is not . The contribution of the simple pole is not damped by the
polynomial.

$\lambda_{N'}$is estimated from a variant of eq. (\ref{eq:2.13})
(with no damping)

\begin{equation}
(2\pi)^{4}\lambda_{N'}^{2}m_{N'}\backsimeq-B_{3}\int_{(m_{_{N}}^{2}+m_{_{N'}}^{2})/2}^{(m_{N'}^{2}+m_{N''}^{2})/2}tdt\label{eq:2.17}
\end{equation}

which yields $\frac{\lambda_{N'}}{\lambda_{N}}\backsimeq.35$

The coupling constant $g_{\pi NN'}$was studied in \cite{Gegelia}
who give $g_{\pi NN'}=.47\pm.04$ . This yields

$\frac{\varDelta_{N'}}{(\text{dominant\,term})}=\frac{g_{\pi NN'}}{g_{\pi NN}}\frac{\lambda_{N'}}{\lambda_{N}}\frac{P(m_{N}^{2})}{P'(m_{N}^{2})}\backsimeq.015$
which shows that $\varDelta$ amounts to only a few percent of the
dominant term. Finally

\begin{equation}
g_{\pi NN}\frac{-P'(m_{N}^{2})}{P(m_{N}^{2})}=\frac{\left[\frac{1}{12\pi^{2}}I_{0}(R)+\frac{4}{3}f_{\pi}^{2}-a_{1}f_{\pi}C\right]}{\left[\frac{1}{4\pi^{2}}I_{1}(R)(1+\frac{3}{2}a_{s})-\frac{1}{12}\left\langle a_{s}GG\right\rangle \right]}\frac{m_{N}}{f_{\pi}}\label{eq:2.18}
\end{equation}

As discussed before the radius $R$ of the integration circle is not
arbitrary, it cannot be too small because this would invalidate the
OPE nor can it be too large because the polynomial would start enhancing
the unknown continuum instead of damping it, $R$ should be chosen
in the stability region of the integrals $I_{0,1}(R)$ shown in Fig.
3

The interval of stability is very wide and the results are 

$I_{0}(R)=.814GeV^{2}$ similarly $I_{1}(R)=.474GeV^{4}$. With the
standard value $\left\langle a_{s}GG\right\rangle =.012GeV^{4}$ eq.
(\ref{eq:2.18}) finally gives

\begin{equation}
g_{\pi NN}=11.5\pm1.0\label{eq:2.19}
\end{equation}

The error, estimated about 8\% stems mostly from neglected radiative
corrections as well as from $\varDelta.$

\begin{figure}

\includegraphics[width=0.6\textwidth]{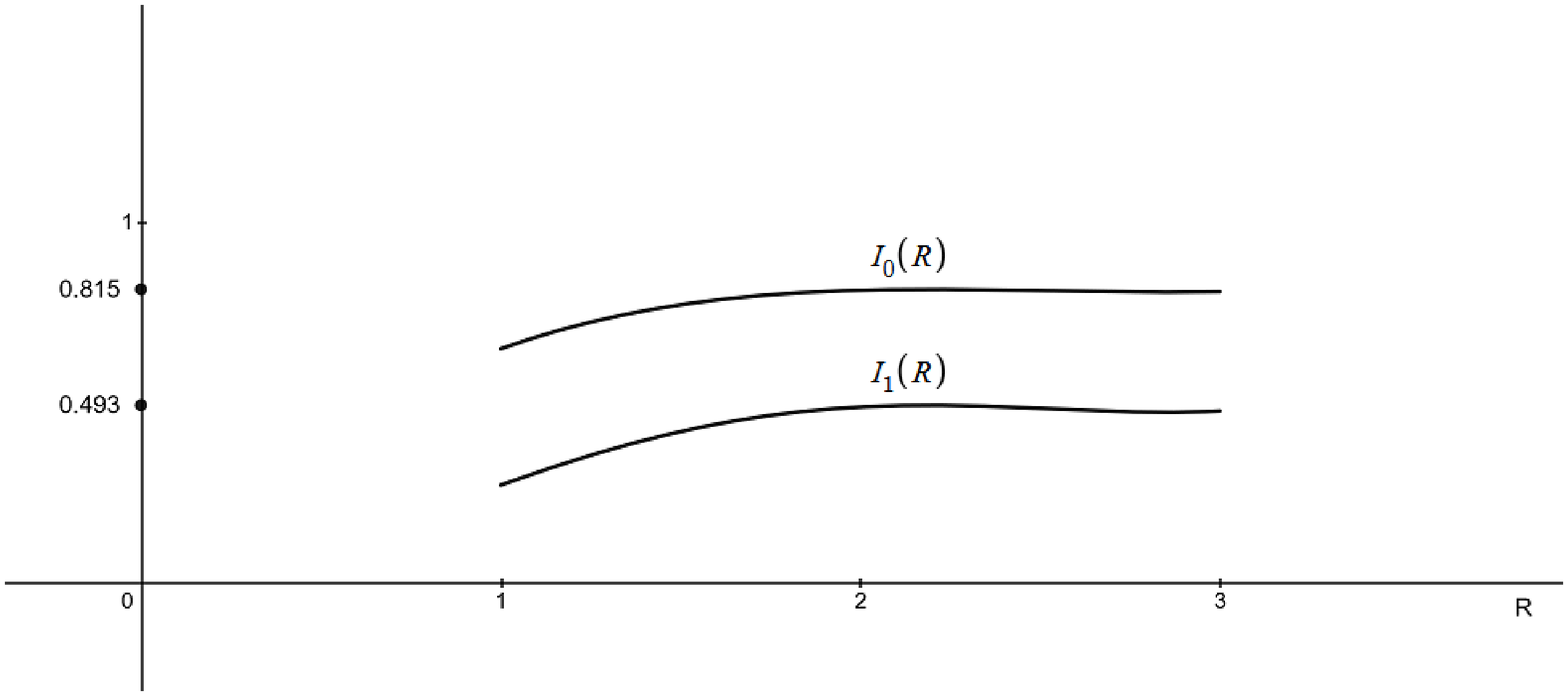}
\centering{}Fig. 3
\end{figure}

In order to assess the stability of the method other choices of the
damping polynomial can be made, e.g $P(t)=(1-\frac{t}{2.5GeV^{2}})$
or impose $\int_{2.0GeV^{2}}^{3.0GeV^{2}}dt\,P(t)=0$. In both cases
the results are very close to eq.(\ref{eq:2.19}) ($g_{\pi NN}=11.2$
and $g_{\pi NN}=11.0$)

\section{Kaon-Baryon Couplings}

Several attempts were made to evaluate the couplings $g_{KN\Lambda}$and
$g_{KN\Sigma}$using QCD sum rules \cite{BNN1},\cite{Choe et al.},\cite{Choe},\cite{BNN2},\cite{Aliev and Savci},\cite{Forkel et al}

which all turned out much smaller than their SU(3) values

\begin{align}
g_{KN\Lambda} & =\frac{-1}{\sqrt{3}}(3-2\alpha)g_{\pi NN}\simeq-13.05\label{eq:3.1}\\
g_{KN\Sigma} & =(2\alpha-1)g_{\pi NN}\simeq3.50\nonumber 
\end{align}

where $\alpha=D/(D+F)\simeq.635$ is the fraction of D-type coupling.

Experimentally, the determination of the coupling constants still
involves uncertainties but the analysis of \cite{Haberzettl} yields

\begin{equation}
g_{KN\varLambda}=-13.5\text{\,\,\,\,\,\,and\,\,\,\,\,\,}g_{KN\Sigma}=4.5\label{eq:3.2}
\end{equation}

Evaluation of the coupling constants using heavy baryon chiral perturbation
theory at the one loop level has also been attempted \cite{Goity}resulting
in values much smaller than one would expect from eq. (\ref{eq:3.1}).

An attempt to calculate the coupling constants starting from the Golberger-Treiman
relation (GTR) in $SU(3)\times(SU(3)$ was undertaken in \cite{NFN}.
In this paper the extrapolation to the nucleon mass shell was done
using exponential kernels to damp the unknown parts of dispersion
integrals. I shall here redo the calculation using polynomial kernels
which avoid the arbitrariness and stability problems.

The GTRs in the strange channel are

\begin{equation}
\sqrt{2}f_{K}g_{KN\Lambda}=(m_{N}+m_{Y})g_{A}^{Y}(0)\label{eq:3.3}
\end{equation}
where $Y=\varLambda,\Sigma.$ g$_{A}^{Y}(0)$ is the axial-vector
coupling constant and $f$$_{K}$ the K-meson decay constant.

The coupling constants could of course be obtained from eq. (\ref{eq:3.3})
because the $g_{_{A}}^{Y}$ are measured. eq. (\ref{eq:3.3}) involves
however an extrapolation in momentum transfer squared from 0 to $m_{K}^{2}$
which is not a small quantity on the hadronic scale. In other words
explicit chiral symmetry breaking leads to corrections to the GTR,
the GT discrepancies (GTD) which are not small in the strangeness
changing case.

The GTD are defined by

\begin{equation}
\varDelta^{Y}=1-(m_{N}+m_{Y})g_{A}^{Y}(0)/(\sqrt{2}f_{K}g_{KNY})\label{eq:3.4}
\end{equation}

The evaluation of the K-nucleon coupling constants will yield the
GTD.

Start from the matrix element

\begin{equation}
\left\langle P\left|\partial_{\mu}A_{\mu}^{K^{+}}\right|Y\right\rangle =\Pi(q^{2})\overline{P}i\gamma_{5}Y\label{eq:3.5}
\end{equation}
q denoting the momentum transfer between the baryons. We have

\begin{equation}
\Pi(0)=(m_{N}+m_{Y})g_{A}^{Y}(0)\label{eq:3.6}
\end{equation}

The analytic properties of $\Pi(t=q^{2})$in he complex t-plane are
known. It has a pole at $t=m_{K}^{2}$ and a cut along the positive
t-axis starting at $t_{th}=(m_{K}+2m_{\pi})^{2}$

\begin{align}
\Pi(t) & =\frac{-\sqrt{2}f_{K}m_{K}^{2}g_{KNY}}{(t-m_{K}^{2})}+\cdots\label{eq:3.7}
\end{align}

Consider now the usual contour $c$ in the complex t-plane and the
integral

\begin{equation}
\frac{1}{2\pi i}\intop_{c}\frac{dt}{t}\Pi(t)=\frac{1}{\pi}\int_{th}^{R}\frac{dt}{t}Im\Pi(t)+\frac{1}{2\pi i}\oint\frac{dt}{t}\Pi^{QCD}(t)=\Pi(0)-\sqrt{2}f_{K}g_{NKY}\label{eq:3.8}
\end{equation}

The first integral on the r.h.s. running along the cut represents
the contribution of the $0^{-}$strange continuum and provides the
main part of the GTD. In the second integral along the circle $\Pi^{QCD}(t)$
is a good approximation for $\Pi(t).$In order to overcome the lack
of knowledge of $Im\Pi(t)$ on the cut consider the modified integral

\begin{align}
 & \frac{1}{2\pi i}\int_{c}dt(\frac{1}{t}-a_{0}-a_{1}t)\Pi(t)\nonumber \\
= & \frac{1}{\pi}\int_{th}^{R}dt(\frac{1}{t}-a_{0}-a_{1}t)Im\Pi(t)+\frac{1}{2\pi i}\oint dt(\frac{1}{t}-a_{0}-a_{1}t)\Pi^{QCD}(t)\label{eq:3.9}\\
= & \Pi(0)-\sqrt{2}f_{K}g_{KNY}(1-a_{0}m_{K}^{2}-a_{1}m_{K}^{4})\nonumber 
\end{align}

The main contribution to the integral over the cut arises from the
interval $I$ : $1.5GeV^{2}\lesssim t\lesssim3.5GeV^{2}$ which includes
the resonances $K(1460)$ and $K(1830)$. The constants $a_{1}$and
$a_{2}$are now chosen so as to annihilate the kernel $(1-a_{1}t-a_{2}t^{2})$
at $t=1.46^{2}GeV^{2}$and at $t=1.83^{2}GeV^{2}$ i.e

\begin{equation}
a_{0}=.77GeV^{-2}\text{\,\,\,\,\,\,and\,\,\,\,\,\,}a_{1}=-.14GeV^{-4}\label{eq:3.10}
\end{equation}

With this choice the integrand is reduced to only a few percent of
its initial value over the interval $I$ and the integral over the
cut becomes negligible. so

\begin{equation}
\Pi(0)-\sqrt{2}f_{K}g_{KNY}(1-a_{1}m_{K}^{2}-a_{2}m_{_{K}}^{4})\backsimeq\frac{1}{2\pi i}\oint dt(\frac{1}{t}-a_{0}-a_{1}t)\Pi^{QCD}(t)\label{eq:3.11}
\end{equation}

It appears from the equation above that chiral symmetry breaking manifests
itself in the presence of the r.h.s. as well as in the deviation of
the factor $(1-a_{1}m_{K}^{2}-a_{2}m_{K}^{4})$ from unity.

In \cite{NFN} $\Pi^{QCD}(t)$ was obtained from Borel-type sum rules.
I shall here redo the calculation using a polynomial kernel as advocated.
Consider the three-point function

\begin{equation}
\varGamma(s=p^{2},t=q^{2})=\int\int dxdye^{-ipx}e^{iqy}\left\langle 0\left|T\psi^{N}(x)\partial_{\mu}A_{\mu}^{K}(y)\psi^{Y}(0)\right|0\right\rangle \label{eq:3.12}
\end{equation}

where $\psi^{N,Y}$are baryonic currents and let

\begin{equation}
\varGamma(s,t)=F(s,t)\sigma_{\mu\nu}\gamma_{5}q_{\mu}p'_{\nu}+\text{other tensor structure}\label{eq:3.13}
\end{equation}

Furthermore the residue at the double baryonic pole of the $\varGamma$
is related to $\Pi(t)$, i.e.

\begin{equation}
\varGamma(s,t)=\left(\frac{\lambda_{N}\lambda_{Y}\Pi(t)}{(s-m_{N}^{2})(s-m_{Y}^{2})}+\cdots\right)\sigma_{\mu\nu}\gamma_{5}p_{\mu}^{'}q_{\nu}+\cdots\label{eq:3.14}
\end{equation}

$\Pi^{QCD}(t)$ is obtained from the above by using $F_{\Lambda}^{QCD}(s,t)$,
extrapolating to the baryons mass shell and identifying terms in (\ref{eq:3.14})
and (\ref{eq:3.15}).

$\lambda_{N,Y}$denotes the coupling of the baryonic currents to the
corresponding baryons.

It will be assumed that the contribution of the single poles is small
as it was in the $\pi-N$ case and it will be neglected.

$\Pi^{QCD}(t)$is obtained by using F$^{QCD}(t)$ and extrapolating
to the baryon mass shell.

$F^{QCD}(s,t)$ is given in \cite{BNN1}

\begin{equation}
F_{\Lambda}^{QCD}(s)=\frac{C_{\Lambda}}{st},\,\,\,\,\,\,C_{\Lambda}=[\frac{4}{3}\sqrt{\frac{2}{3}}m_{s}(\left\langle \overline{q}q^{2}\right\rangle +\left\langle \overline{q}q\overline{s}s\right\rangle )]\label{eq:3.15}
\end{equation}

and 
\begin{equation}
F_{\Sigma}^{QCD}(s,t)\backsimeq0\label{eq:3.16}
\end{equation}

Consider now the integral 
\begin{equation}
\frac{1}{2\pi i}\int_{c}dsP_{Y}(s)F(s,t)\label{eq:3.17}
\end{equation}

in the complex s-plane where as before $P_{Y}(s)$ is the damping
polynomial intended to eliminate the contribution of the continuum. 

This gives

\begin{equation}
\lambda_{N}\lambda_{Y}\Pi^{QCD}(t)\frac{P_{Y}(m_{Y}^{2})-P_{Y}(m_{N}^{2})}{(m_{Y}^{2}-m_{N}^{2})}\backsimeq\frac{1}{2\pi i}\oint dsP_{Y}(s)F^{QCD}(s)\label{eq:3.18}
\end{equation}

and take $P_{Y}(s)=(1-\frac{s}{m_{N'}^{2}})(1-\frac{s}{m_{Y'}^{2}})$
which vanishes at the first nucleon and hyperon excited states ( m$_{N'}^{2}$=1.98GeV$^{2}$,
~~m$_{\varLambda'}^{2}$= 2.40GeV$^{2}$, ~~$\textnormal{m}_{\Sigma'}^{2}=2.51$GeV$^{2}$)

Eqs. (\ref{eq:3.15}) and (\ref{eq:3.16}) inserted in eq. (\ref{eq:3.17})
yield the QCD expressions

\begin{equation}
\frac{\lambda_{N}\lambda_{\Lambda}}{(m_{\Lambda}^{2}-m_{N}^{2})}(P_{_{\Lambda}}(m_{\Lambda}^{2})-P_{_{\Lambda}}(m_{N}^{2}))\Pi^{QCD}(t)\backsimeq\frac{C_{\varLambda}}{t}\label{eq:3.19}
\end{equation}
 and

\begin{equation}
\frac{\lambda_{N}\lambda_{\Sigma}}{(m_{\Sigma}^{2}-m_{N}^{2})}(P_{\Sigma}(m_{\Sigma}^{2})-P_{\Sigma}(m_{N}^{2}))\Pi_{\Sigma}^{QCD}(t)\backsimeq0\label{eq:3.20}
\end{equation}

When this is inserted in eq. (\ref{eq:3.11}) it gives

\begin{equation}
\Pi_{\varLambda}(0)=\sqrt{2}f_{K}g_{KN\Lambda}(1-a_{1}m_{K}^{2}-a_{2}m_{K}^{4})-\frac{a_{1}C_{\Lambda}(m_{_{\varLambda}}^{2}-m_{N}^{2})}{[\lambda_{N}\lambda_{\Lambda}(P_{\Lambda}(m_{\Lambda}^{2})-P_{\Lambda}(m_{N}^{2}))]}\label{eq:3.21}
\end{equation}

\begin{equation}
\Pi_{\Sigma}(0)=\sqrt{2}f_{K}g_{KN\Sigma}(1-a_{1}m_{K}^{2}-a_{2}m_{K}^{4})\label{eq:3.22}
\end{equation}

$\lambda_{N}$is given by eq. (\ref{eq:2.13})

\begin{equation}
(2\pi)^{4}\lambda_{N}^{2}P(m_{N}^{2})=.444GeV^{6}\label{eq:3.23}
\end{equation}

a similar treatment of the $\Lambda$ 2-point function \cite{Choe}
with our polynomial replacing the damping exponential gives

\begin{equation}
(2\pi)^{4}\lambda_{\Lambda}^{2}P_{\Lambda}(m_{\varLambda}^{2})=.481GeV^{6}\label{eq:3.24}
\end{equation}

Finally with the experimental values $g_{\Lambda}^{A}=-.72,\,\,\,\,\,g_{\Sigma}^{A}=.34$
we obtain

\begin{equation}
g_{KN\Sigma}=5.45,\,\,\,\,\,g_{KN\Lambda}=-(11.50\pm1.17\kappa)\label{eq:3.25}
\end{equation}

$\kappa$ denotes the deviation of the 4-quark condensate from factorization,
it arises from the condensate $\left\langle \overline{s}s\overline{q}q\right\rangle \backsimeq\kappa\frac{\left\langle \overline{s}s\right\rangle }{\left\langle \overline{q}q\right\rangle }\left\langle (\overline{q}q)^{2}\right\rangle $
which appears in the expression for the constant C$_{\Lambda}$.

There is no consensus on the value of $\kappa$. I shall take $1\preceq\kappa\preceq3.$

\section{Results and Conclusions}

The $\pi-N$ and K-N coupling constants were calculated using QCD
sum-rules with polynomial kernels tailored to vanish in the baryonic
resonance region. This solves the problems of arbitrariness and stability
inherent to the usual Borel-type QCD sum-rules. The results are

\begin{equation}
g_{\pi NN}=12.4\pm.6,\,\,\,\,\,\,g_{KN\Sigma}=5.45\pm.4,\,\,\,\,\,\,g_{KN\Lambda}=-(12.7-15.0)\label{eq:4.1}
\end{equation}

which are quite close to experiment and to the SU(3) values

The corresponding GTD are large as expected

\begin{equation}
\Delta_{GT}^{\Sigma}\simeq.18,\,\,\,\,\,\Delta_{GT}^{\Lambda}\simeq.29\label{eq:4.2}
\end{equation}

It is finally interesting to see how well our results fit the Dashen-Weinstein
\cite{DW} relation between the GTD

\begin{equation}
g_{\pi NN}\Delta_{GT}^{N}=\frac{m_{\pi}^{2}}{2m_{K}^{2}}(g_{KN\Sigma}\varDelta_{GT}^{\Sigma}-\sqrt{3}g_{KN\Lambda}\varDelta_{GT}^{\Lambda})\label{eq:4.3}
\end{equation}

With $\varDelta_{GT}^{N}\backsimeq.02$ it is seen that relation eq.
(\ref{eq:4.3}) is quite well satisfied.

\pagebreak{}


\begin{thebibliography}{10}
\bibitem{RRY}L. J. Reinders, H. Rubinstein and S. Yazaki, Phys. Rep.
127, 1 (1985) and references therein

\bibitem{SVZ}M. A. Shifman, A. I. Vainshtein and V. I. Zakharov,
Nucl. Phys. B 147, 385, 448 (1979)

\bibitem{S and H}H. Shiomi and T. Hatsuda, Nucl. Phys. A 594, 294
(1995)

\bibitem{B and K}M. C. Birse and B. Krippa. Phys. Lett. B373, 9 (1996),
Phys. Rev. C54, 3240 (1996)

\bibitem{Kim}H. Kim, S. H. Lee and M.Oka, Phys. Lett. B453, 199 (1999)

\bibitem{Azizi}K. Azizi, Y. Sarac and H. Sundu, Eur. Phys. J. A52,
114(2016), hep-ph/arXiv:1510.05432

\bibitem{Joffe}B. Ioffe, Z. Phys C18, 67 (1988)

\bibitem{N S}N. F. Nasrallah and K. Schilcher, Phys. Rev C89, 045202
(2014)

\bibitem{Gegelia}J. Gegelia et al., Phys. Lett. B763, 1 (2016)

\bibitem{BNN1}M. E. Bracco, F. S. Navarra and M. Nielsen, hep-ph/9908452

\bibitem{Choe et al.}S. Choe, M. K. Cheoun and S. H. Lee, Phys. Rev.
C53, 1363 (1996)

\bibitem{Choe}S. Choe, Phys. Rev. C57, 2061 (1998)

\bibitem{BNN2}M. E. Bracco, F. S. Navarra and M. Nielsen, Phys. Lett.
B454, 346 (1999)

\bibitem{Aliev and Savci}T. M. Aliev and M. Savci, hep-ph/9902466

\bibitem{Forkel et al}H. Forkel, M. Nielsen, X. Jin and T. D. Cohen,
Phys. Rev. C50, 3108 (1994)

\bibitem{Haberzettl}H. Haberzettl, C. Bennhold, T. Mart, T. Feuster,
Phys. Rev. C58, 1 (1998)

\bibitem{Goity}J. L. Goity, R. Lewis, M. Schvellinger and L. Zhang,
Phys. Lett. B454, 115 (1999)

\bibitem{NFN}N. F. Nasrallah. Phys. Lett. B647, 262 (2007)

\bibitem{DW}R. Dashen and M. Weinstein, Phys. Rev. 188, 2330 (1969)
\end{thebibliography}
\end{document}